\documentclass{aa}
\usepackage{epsfig}
\begin{document}

\title{TRUFAS, a wavelet based algorithm for the rapid detection of planetary transits}

\author
{C. R\'egulo, 
\inst{1,2}
J. M. Almenara,
\inst{1}
R. Alonso,
\inst{3}
H.J. Deeg,
\inst{1}
\and
T. Roca Cort\'es\inst{1,2}
}

\offprints
{C. R\'egulo}

\institute{Instituto de Astrof\'\i sica de Canarias, 38205, La Laguna, Tenerife, Spain
\and
Dpto. de Astrof\'\i sica, Universidad de La Laguna, La Laguna, 38206, Tenerife, Spain
\and
Laboratoire d'Astrophysique de Marseille, Traverse du Siphon, 13376, Marseille 12, France\\
\email{crr@iac.es}, {jmav@iac.es}, {roi.alonso@oamp.fr}, {hdeeg@iac.es}, {trc@iac.es}
}

\date
{Received ....; accepted ....}

\abstract {} {
We describe a  fast, robust and automatic detection algorithm, TRUFAS, and apply it to data that are being expected from the CoRoT mission.} {The procedure proposed for the detection of planetary transits in light curves  works in two steps: 1) a continuous wavelet transformation of the detrended light curve with posterior selection of the optimum scale for transit detection, and 2) a period search in that selected wavelet transformation. The detrending of the light curves are based on Fourier filtering or  a discrete wavelet transformation. TRUFAS requires the presence of at least 3
transit events in the data.} {The proposed algorithm is shown to identify reliably and quickly the transits that had been included in a standard set of 999 light curves that simulate CoRoT data. Variations in the pre-processing of the light curves and in the selection of the scale of the wavelet transform have only little effect on TRUFAS' results.}  {TRUFAS is a robust and quick transit detection algorithm,
especially well suited for the analysis of very large volumes of data
from space or ground-based experiments, with long enough durations for the
target-planets to produce multiple transit events.}

\authorrunning{ }

\titlerunning{ }

\maketitle

\section{Introduction}

The CoRoT space telescope will observe during its mission  5 or 6 fields for the detection of planetary transits (Rouan et al. \cite{rouan00}, Bord\'e et al. \cite{ borde03}, Garrido \& Deeg \cite{garrido06}, Baglin et al. \cite{baglin06}). Observations on each of these fields will last for 150 days, returning light curves from about 12\ 000 stars with a temporal resolution of 512 s. About 10 further fields will be observed with shorter coverages  of 1 to 2 weeks.
During the observations, frequent searches for the presence of transit-like signatures will be undertaken on the incoming data in order to maximise the scientific return from the mission. If a star shows a promising ``alarm'', its further observing mode can be switched to a higher temporal resolution of 32 s. Also, it is of interest to start complementary ground-based observations of planet-candidates as quickly as possible. Otherwise, since the end of a 150 days observing cycle coincides with the end of a field's visibility,  complementary observations may begin only after several months delay.

When  the satellite is in operation, the transit searches will be repeated weekly on a combination of newly acquired and previous data from the 12\ 000 sample stars. With such amounts of data to be analysed for the existence of planetary transits, fast, robust, and automatic algorithms are vital.  In this paper, we apply a method based on wavelet techniques to the problem of transit detection, that fulfils all these requirements. Part of this method was originally developed for the detection of p-mode oscillations in power spectra of solar-like stars (R\'egulo \& Roca-Cort\'es \cite{regulo1},  \cite{regulo2}). Here, it has been adapted for the detection of transit-like signatures in stellar time-series. The method, though of general nature, is described in this paper in the context of the data expected from the CoRoT mission. This is due to the imminent need of analysis of that mission's data, but also because the availability of a testbench of 999 simulated stellar light curves of CoRoT data. This set had been created for a previous comparison of several detection algorithms within the ``COROT Blind test 1'' (Moutou et al. \cite{moutou}; hereafter BT1), thereby giving us a reference for comparing the algorithm described here. We note that approaches for planet detection through wavelet methods have been published by Jenkins (\cite{jenkins}), who employs a de-noising and matched-filter transit detection in the wavelet domain in the context of the {\it Kepler} mission, and by Husser et al. (\cite{husser}) who gives a preliminary account for a wavelet analysis combined with genetic search algorithms.

The problem of searching for planetary transits in long
duration precision photometry (light curves) of stars, consists of the
search for periodic small brightness-dips in time series. Algorithms, or procedures, that undertake such searches need to deal with a variety of factors. In all known instrumental settings, be they ground or space-based, low-frequency (red noise) and semi-periodic variations are present in the time-series (e.g. Pont et al. \cite{pont06}).
Also, across the brightness range of sample stars there will be varying level signal strengths, and signal to noise
ratios  to deal with. 
The primary parameters defining the transit signatures also vary, with the range of periods to be scrutinized pending on the duration of a field's surveillance. For the case of CoRoT, the periods of interest go from about 1 day to 50-70 days, with the upper limit given by the requirement to detect at least 3 transit events during the 150 days of a field's surveillance, and expected, detectable transit amplitudes range from about 0.04\% for planets with $\sim$ 2  R$_{\oplus}$, to over 1\% from giant planet transits around solar-size stars. The last major parameter that describes a transit's detectability, its duration, varies to a much lesser extent, with $t_{tr} \sim P^{1/3}$, with $P$ being the orbital period. 
Moreover, transit searches are usually made in fields containing several thousands of stars, thereby restricting seriously the computing time that can be spent on one single stellar light curve. 
To maximize the chance of discoveries requires therefore fast, robust and largely automatic  procedures to analyse the light curve of each star. In this paper we propose an algorithm, TRUFAS, that fullfil all these requirements.

\section{Methodology}
The starting point for the transit detection is the observed photometric light curves that have been detrended and filtered by one of the two proposed methods (see section 2.1), either Fourier or Wavelet transformations. The idea of the method is the search for periodicities on a selected wavelet-transformation. This transformation results in a time-series where transit-like signatures are amplified, and the first transit-like signal shows up in the first bin of the FFT of this time-series, independent of the transits' epochs.  The proposed algorithm, TRUFAS, is therefore composed of two principal steps:

\begin{enumerate}

 \item On the detrended data, a Continuous Wavelet Transform is calculated and the scale that shows the largest amplitudes within the lengths of  expected transits is selected.

\item  Periodicities  within the transformed time-series are searched for with the period-searching method developed in R\'egulo and Roca Cort\'es (\cite{regulo1}). This method was designed for the detection of p-mode oscillations in the power spectrum of solar-like stars, based on the fact that the searched peaks are almost equally spaced in frequency. In the case of transits, there will be equally spaced peaks in time, and the same algorithm can be used almost straightforwardly.
\end{enumerate}

An automatic way to reject the false detections from arbitrary noise has also been developed. However, as it will be shown, the algorithm's rate of false  detections is so low, at least when the algorithm is applied to the synthetic light curves for the CoRoT mission, that their individual study do not represent any real problem.

The synthetic light curves are fully described in BT1 in the context of a ``blind'' test, giving an unbiased comparison of transit detection algorithms from several participating teams.
These light curves were generated with an end-to-end instrument simulator of CoRoT (Auvergne et al. \cite{auvergne03}), and include stellar micro-variability (Aigrain et al. \cite{aigrain04}) and an assorted set of planetary transits and transit-like events that were inserted in some of the 999 curves: twenty one contain planetary transits, eleven have low-depth stellar eclipse signals and one results from an eclipsing triple stellar system.

\subsection{Pre-processing the data}

\subsubsection{Detrending and Filtering the light curves}

The CoRoT synthetic light curves are sampled every 512 s during 150 days with quasi-periodic gaps of 30 min each 1.7 h. These gaps simulate the fact that the instrument will cross the South-Atlantic Anomaly (SAA) and these exposures will not be usable.

\begin{figure}[!]		
\psfig{file=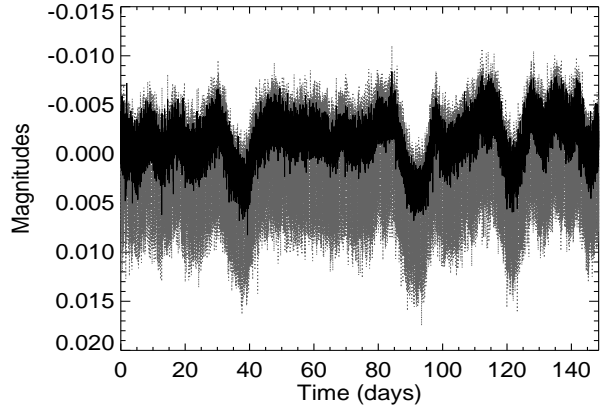,width=8.5cm,height=6.cm}
\caption{Original synthetic light curve  \# 168 generated by the CoRoT team, in grey, together with the same light curve, in black, after removing the earth-scattered residual light and with the gaps interpolated.} 
\label{figorig168}
\end{figure}

The first step in the analysis of the light curves is to remove the  earth-scattered light introduced in the simulated data that varies with the orbital period of the satellite (1.7 hours) and which is not uniform over the CCD. In fact, the contribution that appears in the synthetic  light curves is a residual of this scattered light, that may lead to a positive or negative signal. The effect that introduces this residual earth-scattered light in each synthetic curve, with a period almost following the orbital period, is not exactly the same for all the stars, but it can be inferred directly from each light curve. What we have done in order to remove this effect is to use each light curve to obtain the shape of this spurious signal, identifying a maximum around the orbital period and folding the light curve with the obtained period. Before the folding, each orbit is appropriately set to a common level to avoid the effect of the low frequency noise present in the data that has not yet been removed. The signal obtained after the folding is smoothed and subtracted from the original one. This is done individually for each light curve. After this detrending, the small gaps caused by the crossing of the SAA  are linearly interpolated. The result of this first step is plotted in Fig.~\ref{figorig168}, for the case of the original synthetic light curve \# 168 generated by BT1, which is a light curve containing a transit.  The spiky aspect of the original light curve is due to the orbital 1.7 hours period which is corrected after detrending. However, some low  frequency noise remains that needs to be filtered too, as it is clear from Fig.~\ref{figorig168}.

For the filtering of this ``red noise'', two different types of filters have been tested: a filter in the Fourier domain and a Wavelet filter. This filtering is not an integral part of TRUFAS; but for any light curve based on ``real data'', it is a required previous step. We employ two different filter methods in order to test and verify the reliability of TRUFAS.

\subsubsection{The Fourier Filter}

\begin{figure}[!]		
\psfig{file=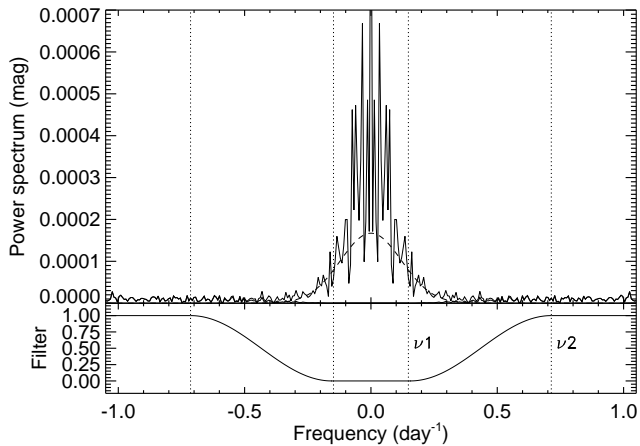,width=8.5cm,height=6.cm}
\caption{Upper panel: Power spectrum of light curve \# 168 with the Gaussian used to select $\nu_{1}$ and $\nu_{2}$.  The lower panel shows the filter function that is being applied to remove the low  frequency noise present in the data. For the definition of $\nu_{1}$ and $\nu_{2}$, see text.}
\label{figpowspec}
\end{figure}

Here, a Fast Fourier Transform (FFT) is performed on each detrended light curve, and its power spectrum calculated. The low-frequency domain of the spectrum is then fitted by a normalized Gaussian function (upper panel in Fig.~\ref{figpowspec}). The Gaussian is only used to define two frequencies  $\nu_{1}$ and $\nu_{2}$ of a filter-function, which are the frequencies where the Gaussian has values of 0.5 and 10$^{-8}$ respectively, which implies that $\nu_{2} = 5.155 \nu_{1}$. 
The filter function (lower panel in Fig.~\ref{figpowspec}) has zero value between zero frequency and $\nu_{1}$ and is 1 above $\nu_{2}$. The change from 0 to 1 between $\nu_{1}$ and $\nu_{2}$ is smoothed using a half sine function in order to avoid the Gibbs effect. After multiplying the power spectrum with the filter function, an inverse FFT procedure generates the final light curve. The result of this filter is shown in Fig. 3 (upper panel) for light curve \# 168 of the synthetic data set.

\begin{figure}[!]		
\psfig{file=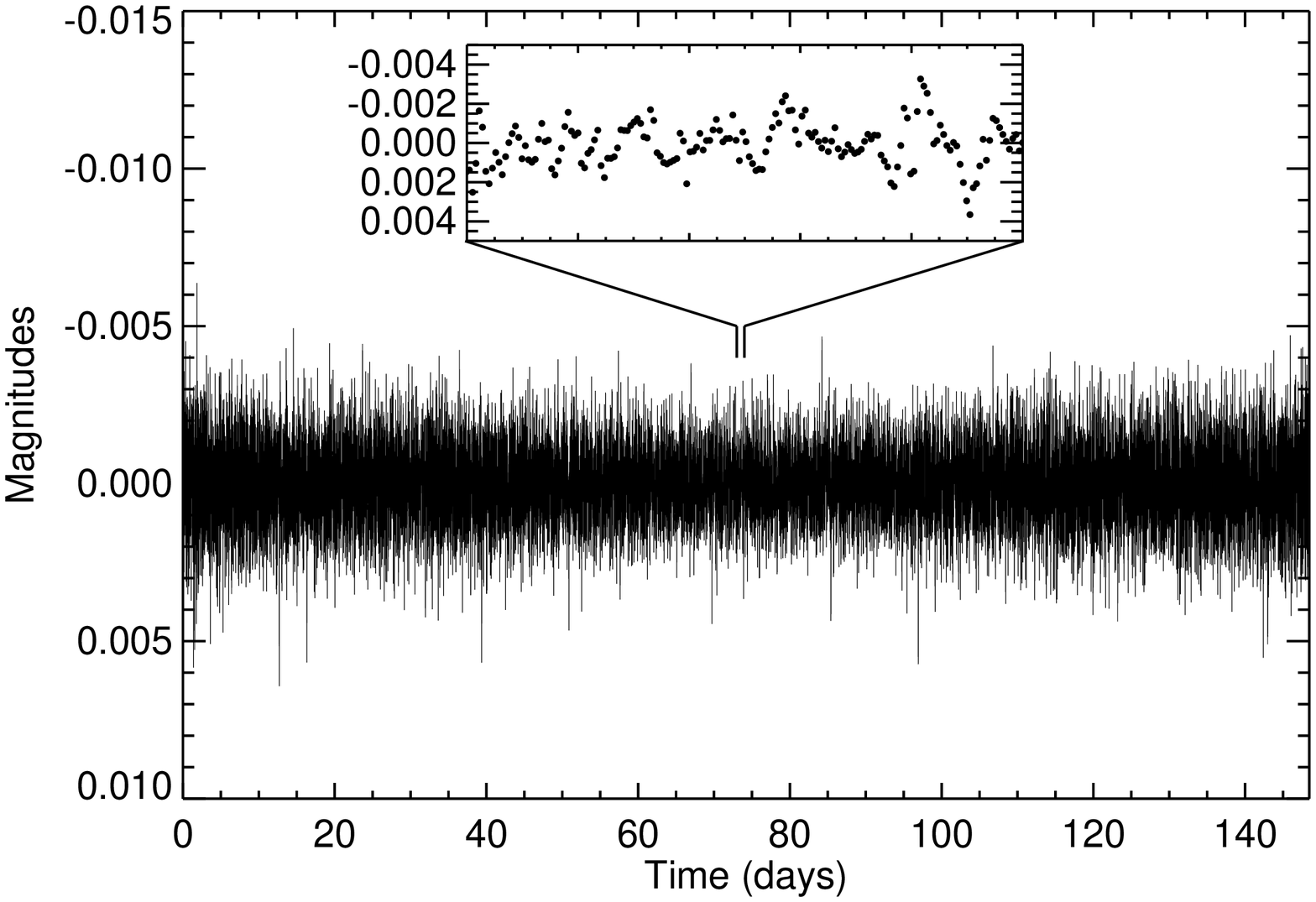,width=8.5cm,height=6.cm}

\psfig{file=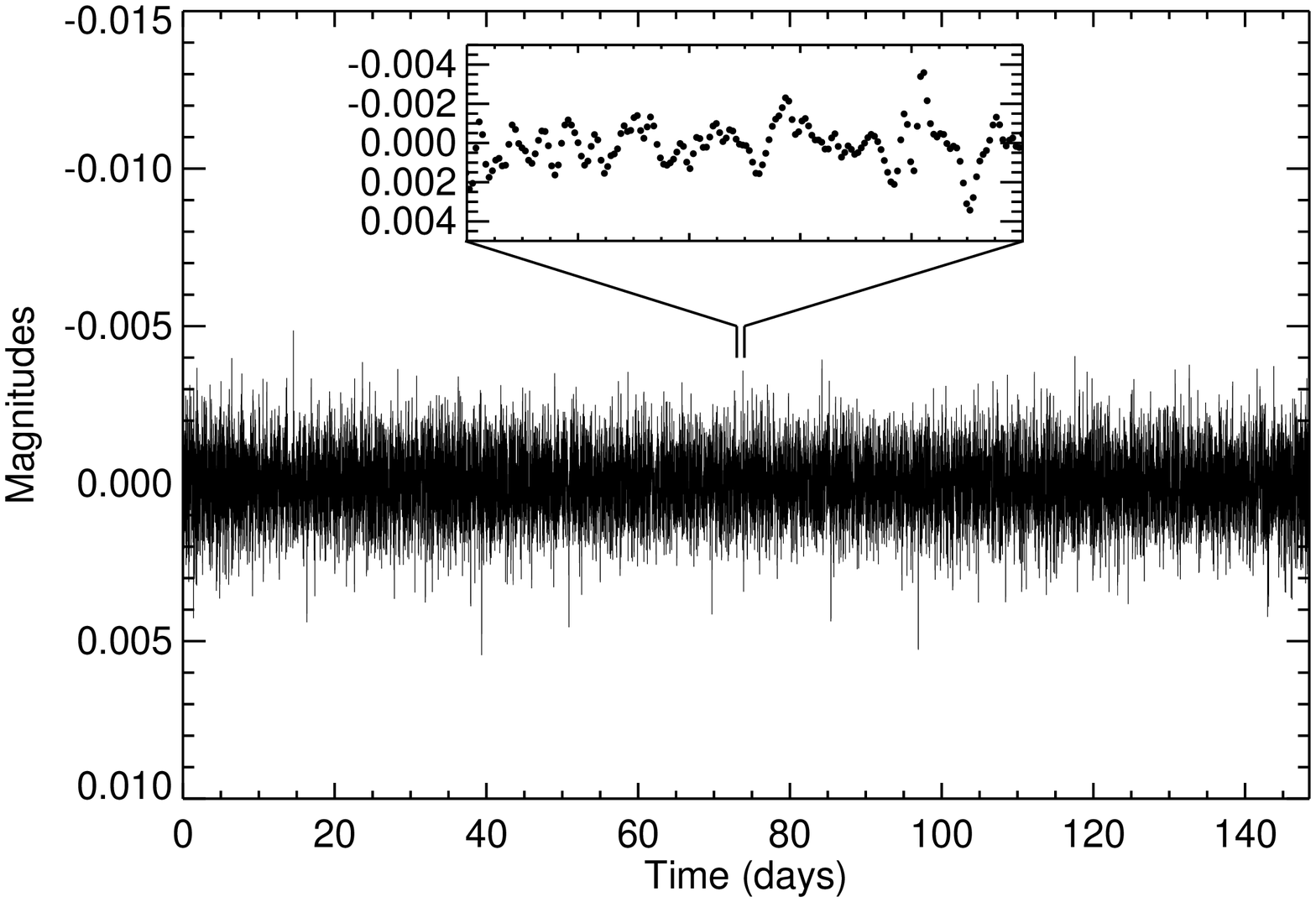,width=8.5cm,height=6.cm}
\caption{Filtered light curve \# 168. Upper plot, with a filter using FFT. Lower plot, from filtering using DWT. The amplifications in the inserts show the data over one day.}
\end{figure}
 \label{figfilt}
 
 \subsubsection{Filtering with Wavelets}

Wavelet techniques provide a method to de-noise signals only  at the time that the noise occurs, and without significant degradation of the signal, as opposed to conventional filtering which remove noise across the whole signal  (Daubechies \cite{daub}).

In a Wavelet Transform (WT), the signal is decomposed into a set of basic functions called wavelets that are spatially localized. These wavelets are obtained from a single one, known as a ``mother'' wavelet by dilations and contractions (scaling) as well as shifts. Therefore, in a WT, the concept of scale is introduced as an alternative to frequency.

The WT is called continuous (CWT) when it scales and shifts the mother or base wavelet through a continuous range of values; this kind of CWT will be used in later steps. The WT is discrete (DWT) when it scales and shifts the base wavelet only in powers of 2. Such a DWT is used here to filter the signal.

The DWT analysis allows us to retrieve two different aspects of a signal: approximation coefficients (ApC) and detail   coefficients (DtC). ApC are the low frequency components of the signal and DtC are its high frequency components, usually related with noise. If a signal has, for instance, 1024 points, each set of coefficients, ApC  and DtC, will have 512 points. This step  can be repeated over the ApC  to obtain a second level approximation and detail coefficients  of 256 samples each (Young  \cite{young}). We can continue repeating the sequence until a DWT at level 1 yields wavelet coefficient at a scale $2^{1}$. The filter is performed removing the set of coefficients where the noise lies and using the inverse discrete wavelet transform to reconstruct the de-noised synthesized signal.

In our case, the signal is separated into 14 scales, corresponding to wavelets of different width, each one double the previous one.  We multiply by 0 or 1 each scale before applying the inverse transformation to obtain the de-noised light curve. The removed scales are the first scale of details, that correspond with high frequencies, as well as the last 6, 7 or 8 scales of approximations that correspond with low frequencies. How many ApC scales are multiplied by zero depends on the signal. The selection is performed automatically by measuring the dispersion of the light curves; the higher the dispersion the stronger the filter, that is, more scales are removed.

The function used as a mother wavelet was a Daubechies order 24 (Daubechies \cite{daub}) . Although many different functions that produce different coefficients can be used as a mother wavelet, for our filtering process the final result is almost independent of the used function. The result of  this filter is shown in Fig. 3 (bottom) for the light curve \# 168. 

\subsection{ TRUFAS Algorithm}

\subsubsection{ TRUFAS' First Step: A Continuous Wavelet Transform \label{sec:scale}}

 TRUFAS starts with a Continuous Wavelet Transformation (CWT) (Torrence and Compo \cite{torrence}) of the
filtered data, which is used to select the part of the signal where the transit search will be done.

\begin{figure}[!]		
\psfig{file=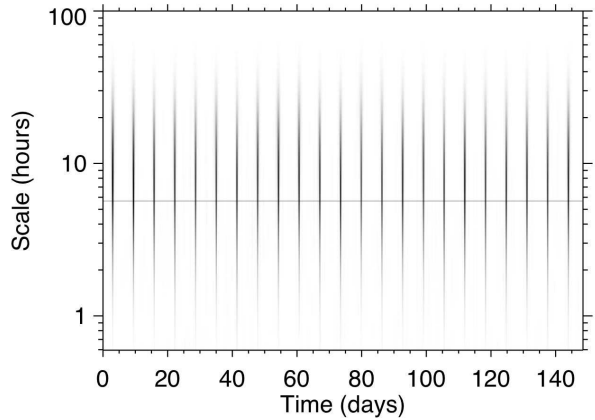,width=8.5cm,height=6.cm}
\caption{CWT of light curve \# 533 (light curve with clear transit signal) that maps the correlation coefficients of 55 different scales, ordered by their corresponding transit-duration in hours. The horizontal line corresponds with a transit duration of 5.7 h.}
\label{fig55scales}
\end{figure}

\begin{figure}[!]		
\psfig{file=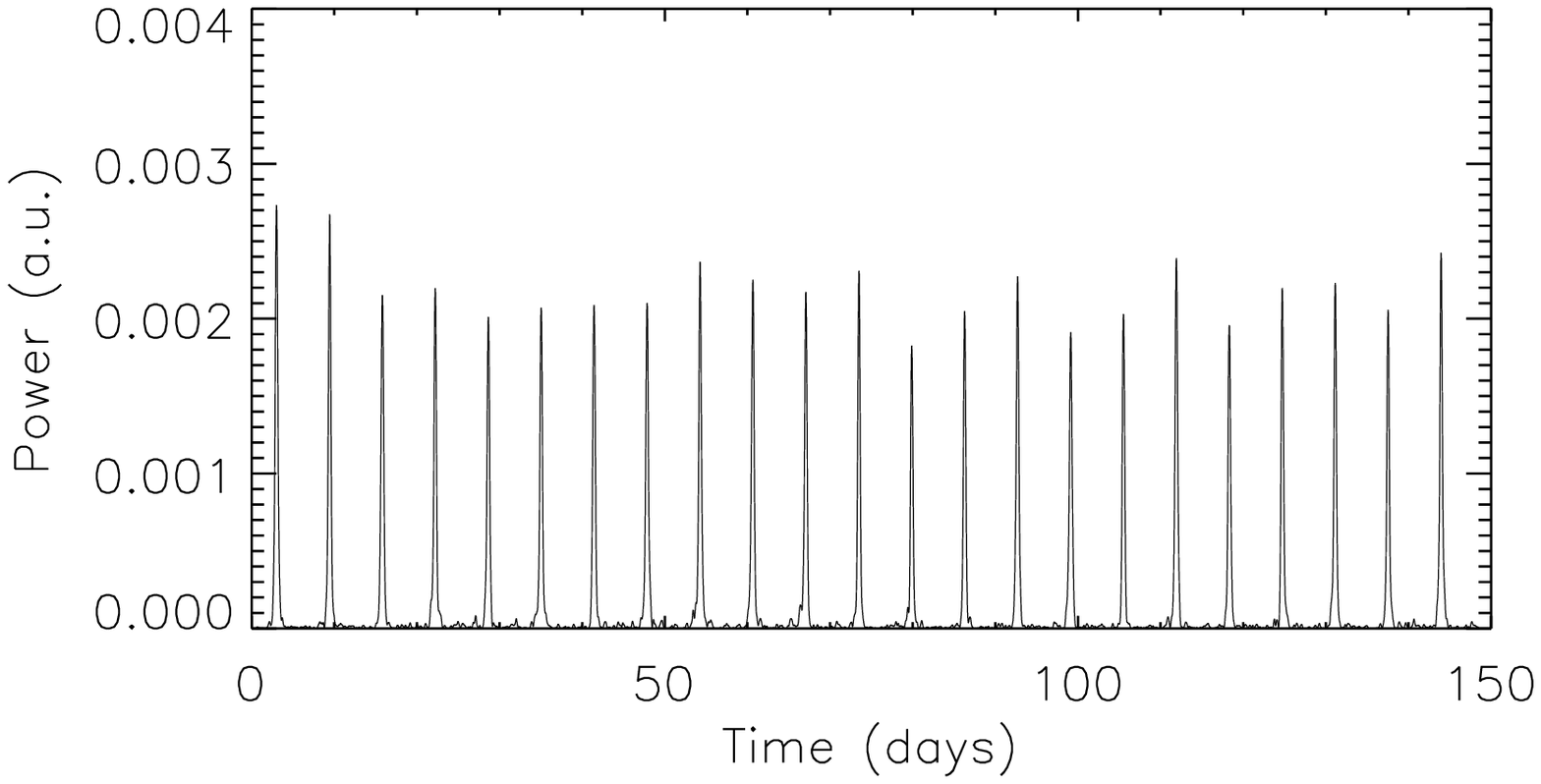,width=8.5cm,height=6.cm}

\psfig{file=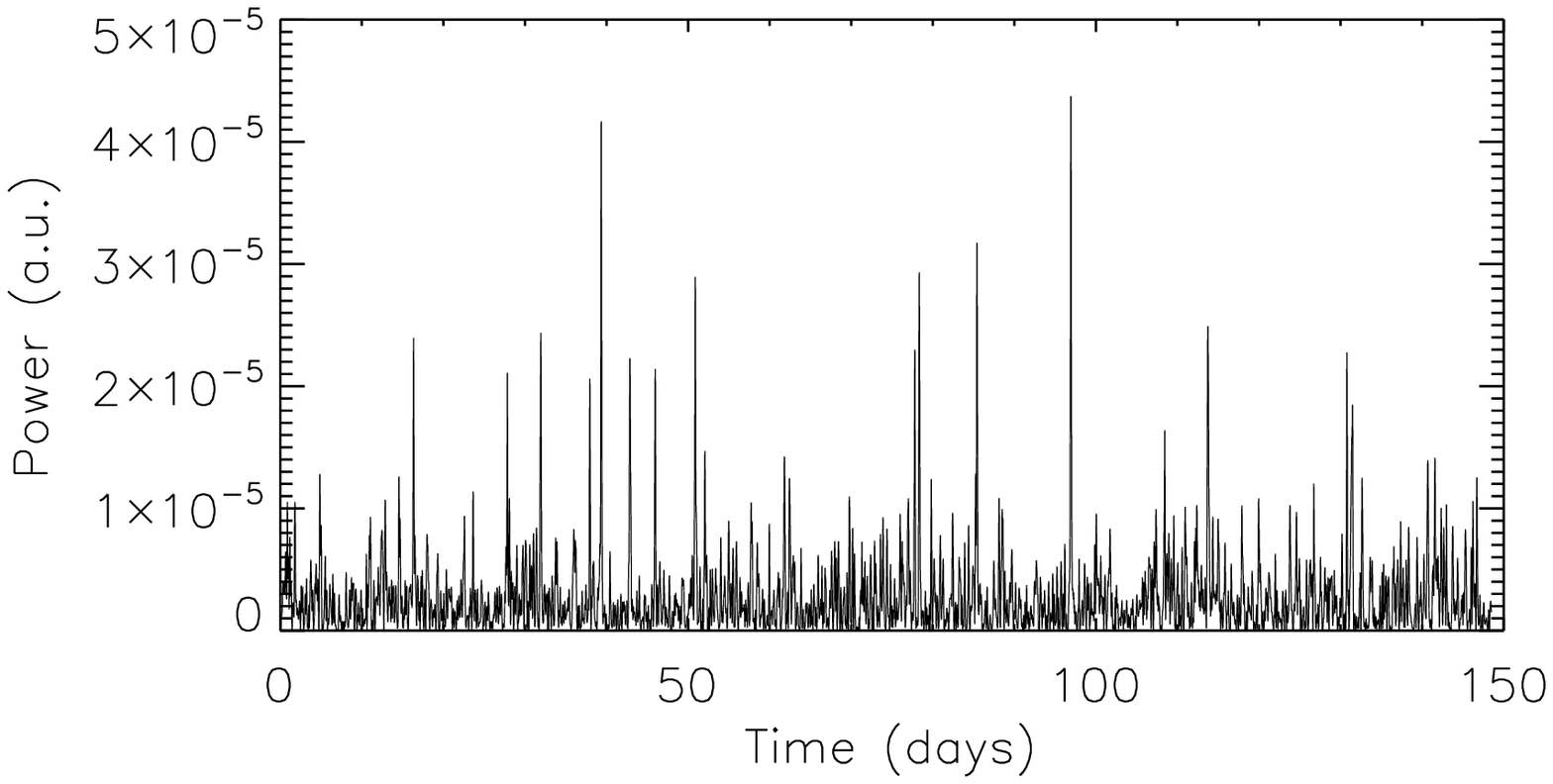,width=8.5cm,height=6.cm}
\caption{Two examples for the selected scaled of two light curves.  The units of the vertical axis (``power'''') are the square of the signal of the CWT. Top panel: light curve \# 533. Bottom panel: light curve \# 168.}
\label{figselscale}
\end{figure}

The continuous wavelet transform of a function $f({\eta})$ is defined by:

\begin{displaymath}
WT = \int{f(\eta)\, \; \Psi^{*} (\eta) d \eta}.
\end{displaymath}
Our ``mother'' wavelet was a Paul function (Torrence and Compo \cite{torrence}) :

\begin{displaymath}
 \Psi( \eta) = \frac{ 2^{m}\, \;  i^{m}} {\surd \pi \, (2m)!}  \, \; (1 - i \eta)^{-(m + 1)} \;  \;  \;  \; with \, \;  \;
 m = 1,
\end{displaymath}
and the scaled wavelet is :
\begin{displaymath}
\Psi ( \frac{ \eta - n} {s}) = (\frac {1} {s})^{1/2} \, \;   \Psi_{o} ( \frac{ \eta - n} {s}),
\end{displaymath}
where  $\eta$ is the time in this case, $s$ the dilation parameter used to change the scale, and $n$
the translation parameter used to slide in time. The factor of
$s^{-1/2}$ is a normalisation factor to keep the total
energy of the scaled wavelet constant. Therefore, the CWT maps the signal in a two dimension function in a time-scale space. 
The decomposition was made in 55 scales. The Paul function order 1 was selected because its shape is similar to the feature we are looking for. In fact, the continuous wavelet transform is just a correlation between the function, the light curve in our case, and the scaled and shifted wavelet function. The higher the correlation, the higher the coefficient of the CWT. From the 55 used scales, see Fig.~\ref{fig55scales}, an automatic selection of the best scale was performed . The selection of the scale was done following the double criteria that the best scale is the one with higher coefficients  when these high coefficients are present in more scales. A scale may be considered to correspond to a transit with a given duration if the width of the central, Gaussian-like, part of the scale has a duration similar to that transit. When the automatic scale-selection gave results outside the range corresponding to transits, between 2.3 and 9.5 hours (in the BT1 set of test light curves, one may only expect durations in that range), scales that correspond to transit durations of 5.7 h were used, being a good compromise for the expected lenght of transits. Such failures of the automatic scale selection happen in cases of transits of very low S/N. 
The use of scales significantly deviating from the expected transit duration could introduce problems in the next step of the algorithm: when the scales are too narrow, peaks related to high frequency noise could produce false detections in the transit-search algorithm; on the other hand, scales that are too wide produce signals that are also too wide to be a suitable input to the peak-searching algorithm that is used in the next step.
Fig.~\ref{figselscale} shows two examples of selected scales. As it was pointed out previously, the selected scale is a one dimensional
function of the correlation coefficient between the
light curve and the wavelet transform, versus time. When a planetary transit with high S/N is present in the data (Fig.~\ref{figselscale}, top), it appears as equally spaced peaks in the selected scale. Transits whose amplitude is comparable to the noise will lead to a set of peaks (Fig.~\ref{figselscale}, bottom), some from the transits and some from noise, without apparent  periodicity. The search for periodicities among these peaks is done  in the next step, using the power (the square of the signal) of the selected scale.  

\begin{figure}[!]		
\psfig{file=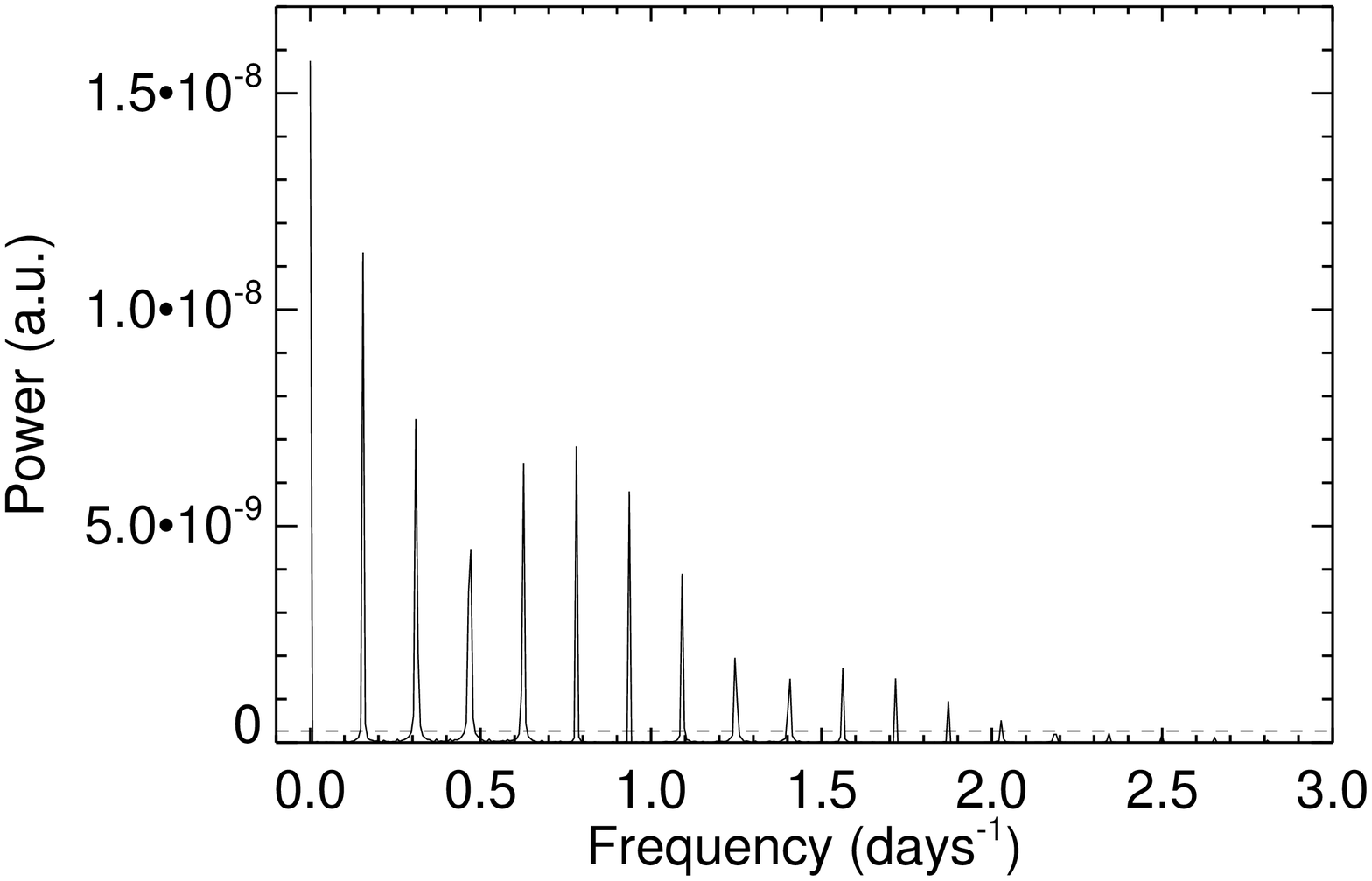,width=8.5cm,height=6.cm}

\psfig{file=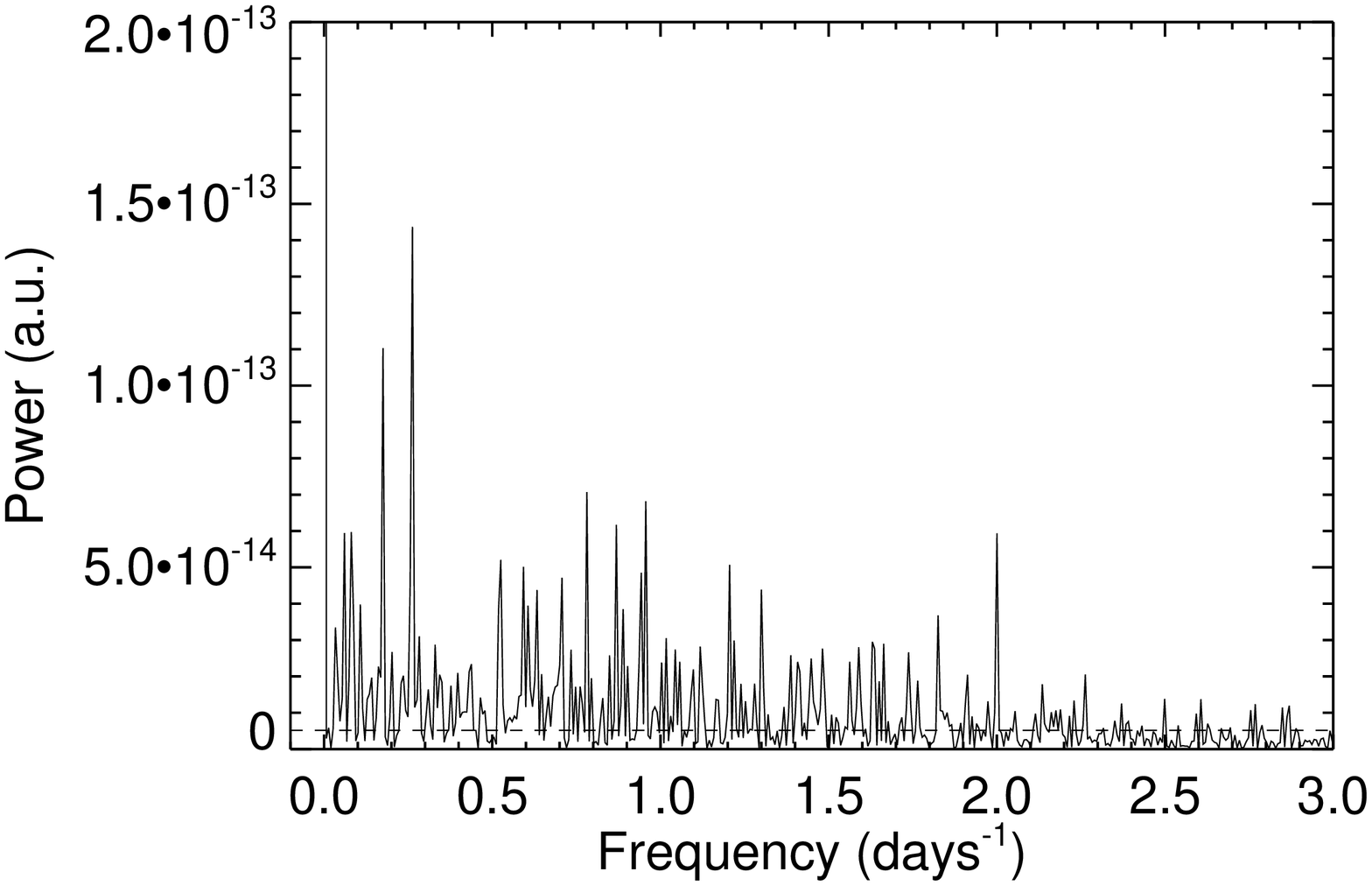,width=8.5cm,height=6.cm}
\caption{FFT of selected scale for light curves \# 533 (Top) and \# 168 (Bottom).  In both graphs, the horizontally dashed line near the zero-level shows the threshold above which peaks are selected for the search for planetary transits.}
\label{figFFT}
\end{figure}

\begin{figure}[!]		
\psfig{file=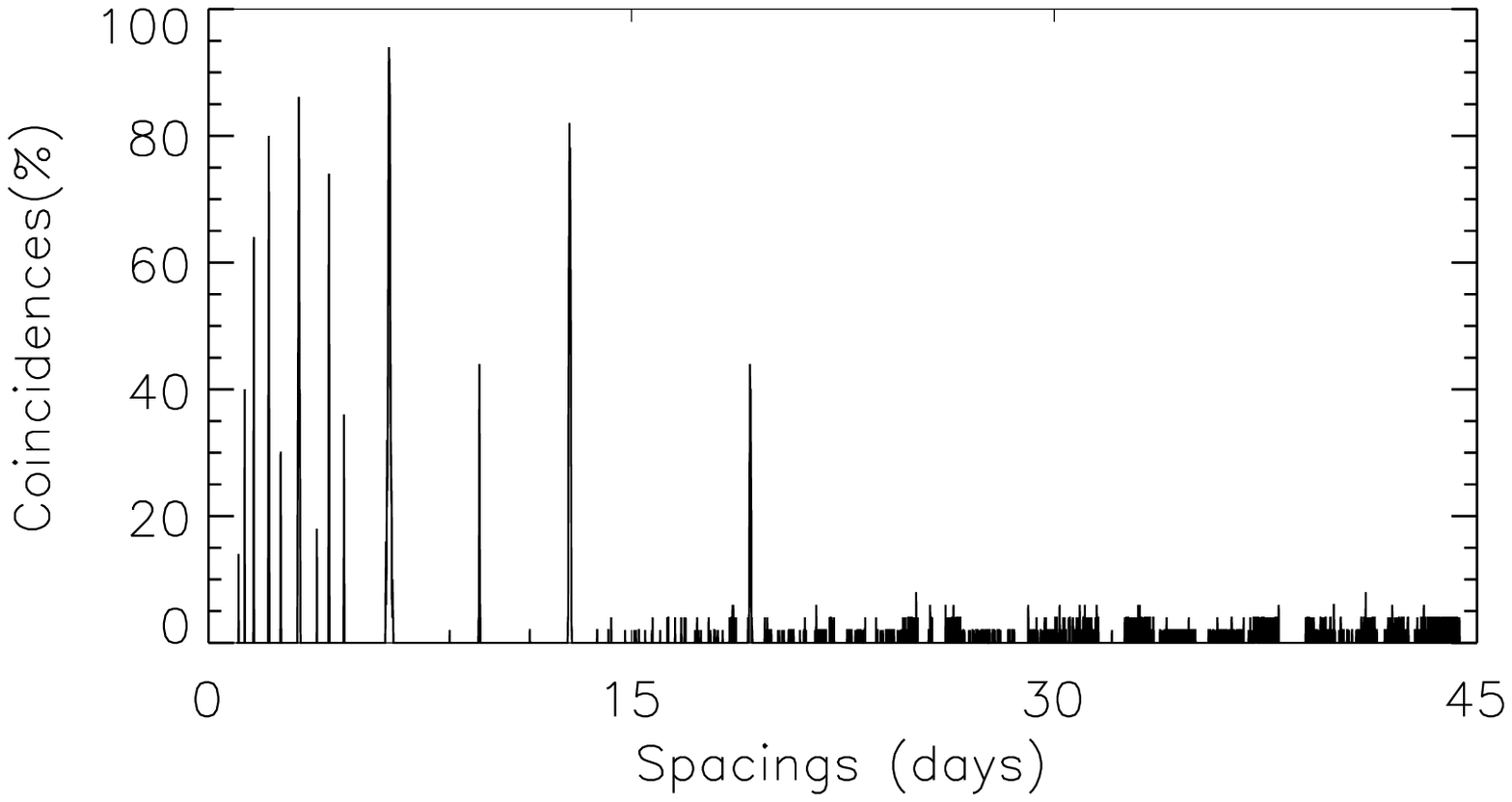,width=8.5cm,height=6.cm}

\psfig{file=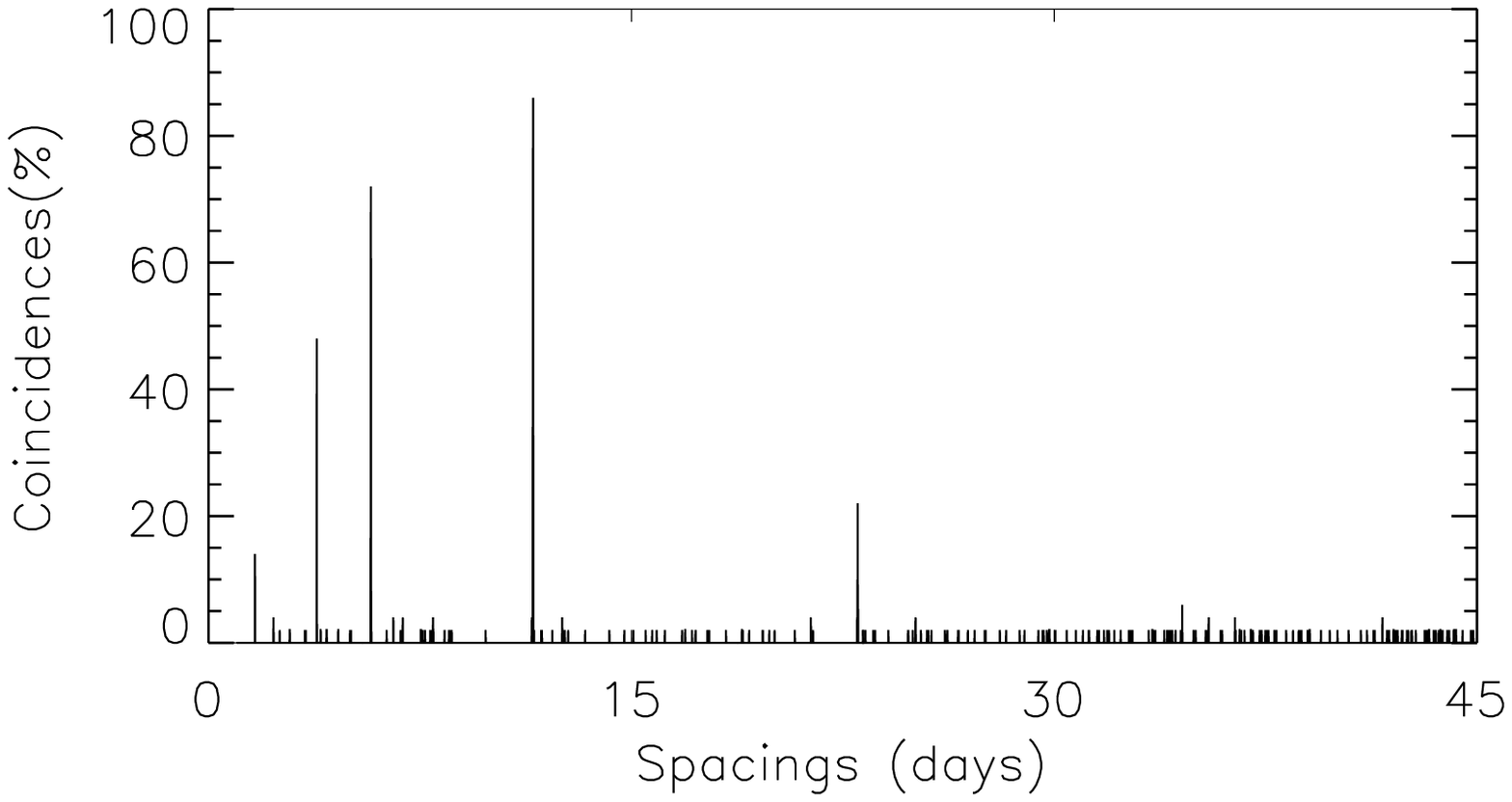,width=8.5cm,height=6.cm}
\caption{Results of the search for the spacing in the FFT of the selected scale on light curve \# 533 (Top) and  \# 168 (Bottom). In both cases, the highest peak corresponds to the period of the transit present in the data. The rest of the peaks are multiples and submultiples of the period present in the data.}
\label{figspacing}
\end{figure}

\subsubsection{TRUFAS' second step: Finding regularly spaced peaks in light curves with transits}

If a light curve contains  transits in it, there will be a set
of peaks regularly spaced in time by an amount that coincides with
the period of the planet. It is this feature or pattern that defines
the signature of the transit. The level of the
amplitude of the transit relative to the noise present in the data
 will set the difficulty in automatically selecting which light
curve contains a true transit.

To find this signature the method developed in R\'egulo and Roca Cort\'es
(\cite{regulo1}), has been applied. What the method detects is the
spacing among the equidistant set of peaks.\\

In brief, the method works as follows. The starting point is the square of selected scale from the previous step, containing the equally spaced peaks we intend to find. The next step is to obtain the power spectrum of this signal by performing a FFT. This spectrum is again a series of equally spaced peaks, but now the first peak is at zero frequency, independent of the epoch or phase of the transits (see Fig.~\ref{figFFT} for the FFT of the light curves shown in Fig.~\ref{figselscale}). Finding the spacing among the peaks is now much easier,  knowing the position of the first one. The search for the spacing  ($T$) is done iteratively trying a range of values that in our case covers from 1 day to 60 days in steps of 512 s, which corresponds to the temporal resolution of CoRoT data. We try to find if there is a signal 1.5 times above the rms of the power spectrum at any of the bins spaced $\nu_{0}$ = $T^{-1}$ apart. To evaluate the significance of the found peaks and to avoid binning effects, this procedure is repeated 50 times on the selected scale, but continuously shortening its length, until it is shortened for about 10$\%$.  The coincidence of  periods among peaks found in the 50 trials is then registered. This method of searching for the period also finds the multiples and submultiples of any periodicity present in the data.

TRUFAS was applied to the 999 synthetic light curves of BT1, and Fig.~\ref{figspacing} shows typical results of the transit search. In the example in Fig.~\ref{figselscale}, light curve \# 533 has a very clear planetary transit, and the period found by  TRUFAS is 6.3985 days with a level of coincidence of 94 \%.  Light curve \# 168 has a weak transit signal that was found by only 3 of the 5 algorithms compared in BT1. Here TRUFAS found a signal at a level of coincidence of 86 \% among the 50 trials, with a periodicity of 11.5125 days. The other peaks that appear in Fig.~\ref{figspacing} are the multiples and submultiples of the found period.

\subsection{Rejecting False Detections\label{sec:falsedet}}

In the automatic application of the algorithm to the 999 CoRoT synthetic light curves, a threshold of the coincidence value was set above which peaks are selected as real transits. As a consequence,  some false detections may appear. By false detection we mean any detection that does not correspond to any transit-like feature with astrophysical origin, being it a planet or some other stellar configuration simulating one. False detections in this sense are therefore detections caused by some random noise. The number of false detections depends on how well the noise has been filtered and on the level of the threshold for the coincidence on the trials' results. With well chosen thresholds (see section \ref{sec:disc} for the selection of thresholds), the rate of false detections from TRUFAS turns out to be less than 1 \%.

\begin{figure}[!]		

\psfig{file=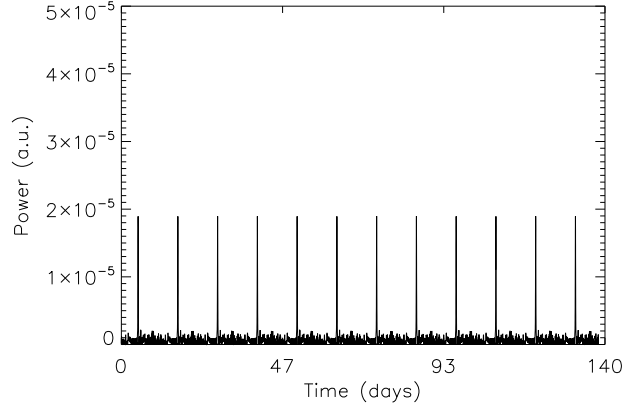,width=8.5cm,height=6.cm}
\caption{Recovered transit events for light curve \# 168. To be compared with  the lower panel of Fig.~\ref{figselscale}.}
\label{figrecovered}
\end{figure}

The rejection of false detections is based on the reconstruction of the selected scale based on the knowledge of the spacing of the peaks, by selecting in the signal's complex Fourier Transform only those bins spaced $\nu_{0}$ apart, and performing an inverse Fourier Transform.  The result is a recovered signal with much better S/N, as it can be seen in  Fig.~\ref{figrecovered}. This recovered signal can be used for an automatic rejection of false detections. When the amplitude (A) of the recovered signal is compared with the sigma ($\sigma$) of the selected scale,  the ratio A/$\sigma$  defines a threshold of higher than 1 for real transits and less than 1 for false detections.  This is due to the fact that during the signal recovery, if the selected bins are not generated by a signal present in the data, their phases do not have the correct relationship to reconstruct the signal and only noise appears. In that case, the amplitude of the recovered signal is at the level of the recovered noise. The recovered noise is lower than the noise present in the original signal, because a significant percentage of bins has been set to zero before recovering the signal; hence the ratio A/$\sigma$ will be lower than 1.
For instance, in the two cases shown, stars \# 168 and \# 533, these ratios are 4.58 and 4.73  respectively whereas all the false  detections with level of coincidences above 20 $\%$ have values of A/$\sigma$ between 0.39 and 0.80.  It is important to keep in mind here, that $\sigma$ and A are not obtained from the same signal but rather from the selected scale before and after signal recovery.

\subsection{TRUFAS in ground-based data: Light curves with gaps}

Although TRUFAS has been tailored to space observations (data without or with rather few gaps, that can be easily interpolated),  in the following we show how the algorithm performs in data with gaps. Out of each of the two light curves that have previously been used as examples, \# 168 and \# 533, we have generated two sets of 100 curves with different duty-cycles by randomly introducing gaps on a 24 hours basis interval.  The gaps were uniformly distributed. These  sets contain curves with duty cycles ranging from 86 \% to only 16 \%, and hence include duty cycles that are typical for ground based observations. 

\begin{figure}[!]		
\psfig{file=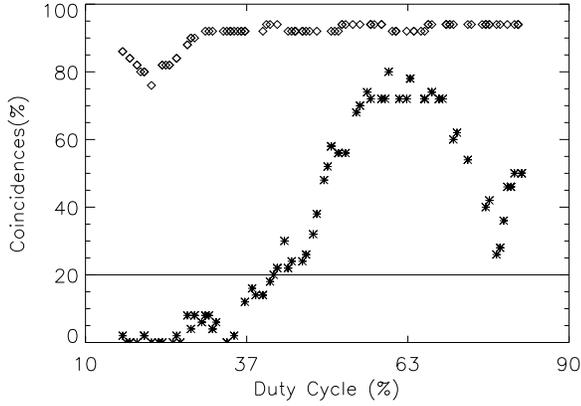,width=8.5cm,height=6.cm}
\caption{Level of coincidences found in  the search for the spacing in the FFT of the selected scale on light curve \# 533 (diamond) and  \# 168 (asterisk)  vs the duty cycle, in 100 different realizations with gaps for each light curve. The line shows the threshold for a confidence detection of a transit (see section 3).}
\label{figgaps}
\end{figure}

The results of the analysis are shown in Fig.~\ref{figgaps}, where the output from TRUFAS (coincidences) is plotted versus the duty cycle. It is interesting to notice that in both cases, the relation between the duty cycle and the  level of coincidences is not linear. In fact this is not  unexpected because a positive detection depends on the relative timing between transits and  gaps. Due to this relative timing it is possible to have bad results with high duty cycles, if many transits happen to be in gaps. More specifically, for light curve \# 533, (with high S/N), the results are very good, because even with a duty cycle as small as 16 \%, the level of coincidences is as high as 75 \%, yielding clear detections. In the noisier light curve \# 168, the results are quite good too, with duty cycles above 40\% leading to detections, if a minimum level of coincidences of 20\% is required (see discussion to justify this number as a good threshold). The decrease in coincidences at duty cycles between 63\% and 80\% is due to the stroboscopic effect: as twice the period of \# 168 is close to an entire number (30 d), for short duty cycles only one of every two transits is present in the light curve, producing transits with twice the real period. As the duty cycle increases, the light curve contains events separated either by T or 2T, what makes more difficult to find a periodicity.  A similar effect can be seen in Fig. 1 of O'Donovan et al. (\cite{donovan}).

In both cases we have computed the coincidences for the bin that correspond to the period of the transit in each light curve, 6.3985 days for light curve \# 533 and 11.5125 days for \# 168. As the method detects not only the period but its  multiples and submultiples it is possible to have higher number of coincidences for any of these multiples or  submultiples, something that can help to find the transit in the difficult cases. This has not been taken into account in this exercise.

\section{Results and Discussion  \label{sec:disc} }

\begin{table}[!bp]
\caption{Number of detected transits in four different configurations of TRUFAS and in dependence on the required level of coincidence (LC). The numbers of false detections are given in brackets.  FF = Fourier Filter, WF = Wavelet Filter, AS = Auto-selected Scale, FS = Fixed Scale.}
\begin{center}
\begin{tabular}{c c c c c} \hline 
 LC (\%) & FF-AS & WF-AS    & FF-FS   &   WF-FS      \\  \hline
 $>$10 & 23 (50) &  22 (35) & 23 (28) &  22 (32)     \\
 $>$20 & 23 (9)  &  22 (5)  & 23 (7)  &  22 (6)     \\
 $>$30 & 23 (5)  &  22 (2)  & 23 (2)  &  21 (3)     \\
 $>$40 & 23 (2)  &  22 (1)  & 22 (1)  &  21 (2)     \\ \
 $>$50 & 22 (0)  &  21 (1)  & 21 (0)  &  20 (1)     \\ \hline
 
\end{tabular}
\end{center}
\end{table}

\begin{figure}[!]		
\psfig{file=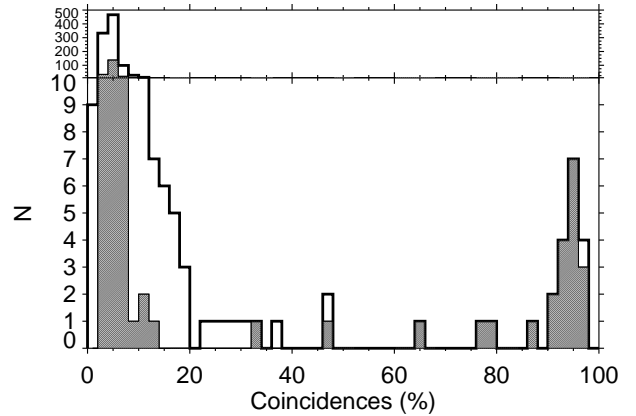,width=8.5cm,height=6.cm}
\caption{Histogram from the analysis of the 999 trial light curves, indicating the coincidences of the detected periods across the 50 trials that were performed on each light curve. The greyed area shows the subset for which the ratio of the recovered amplitude in the filtered scale over the noise in the original scale, A/ $\sigma$, is larger than 1. In combination with high coincidences (10-50 \% was used as threshold, see Table 1), these cases constitute the planet candidates. On the contrary, cases with high coincidences and A/ $\sigma < 1 $ are false detections. Note the different scale for N $> 10$.}
\label{fighistcoin}
\end{figure}

To analyse the behaviour of TRUFAS, on the 999 synthetic light curves we try four different combinations with very similar results. First, we filter the raw data either with the Fourier Filter or with the DWT. Second, for the period search, either an  automatically selected scale (as explained in section \ref{sec:scale}), or a fixed scale corresponding to a transit duration of around 5.7 hours are used; the latter one being in the middle of the expected range of transit durations. Another variable is the threshold for the required fraction of coincidences across the 50 repetitions of one TRUFAS run. The results for the four combinations and different  thresholds of coincidences (10, 20, 30, 40, 50\%) are shown in Table 1. Also, a histogram of the coincidences in the 999 lightcurves, for the combination Fourier-Filter and fixed scale, is shown in Fig~\ref{fighistcoin}.
In Table 1 we note a maximum number of detected transits of 23.  A threshold of the coincidence of 10 \% is too low, because the number of false detection is high while the number of transits detection is comparable to the results when using higher thresholds. Minimum coincidences of 20 -30 \% may be  useful values; they produce less then 1 \% false detections (out if the 999 input light curves) while still recovering  the same number of real transits. Coincidence thresholds above 30\% have very low false detection rates but may lead to the loss of some real transits. In real applications, with many times more input light curves to be processed; these might however still be useful to select more reliable candidates only.

When the four different combinations of filtering and scale-selection are compared, no important differences appear on detection rates; we note only that with the Fourier Filter one more transit is recovered that with the DWT filter. There is no consistent dependence between false detections and filtering method; whereas somewhat fewer false detections are obtained with the fixed scale, apparently a result from its lower sensitivity to events that do not have duration on the order of 5-6 hours. In fact, all the combinations give very similar results, showing that they all are good enough to be used, being one slightly better than the other depending on the noise present in the data and the events that are being searched for. 
We note here that a possibility to optimize the sensitivity to real transits is to limit the automatic scale-selection to transit-durations that may be expected from the combination of a light curve's known stellar parameters (mass, radius) and the period that is being searched (e.g. different scale-selections may be used for different ranges of period-searches).

\begin{table}[!bp]
\caption{Comparison of detections of planets and eclipsing binaries in BT1 and TRUFAS. Flags for detections (+) and misses (--) are given for the five teams of BT1 (see text), and for TRUFAS, together with the period identified by TRUFAS (in the second column are the simulated periods for comparison). The lightcurves with IDs in italics have  period above 50 days which was outside the search range of TRUFAS. In the upper part (PT) are the 21 planetary transits present in the 999 generated light curves and in the lower part (EB) are the 11 eclipsing binaries and the triple stellar system (ID 249) that had been included.}
\begin{center}
\begin{tabular}{c c c c c} \hline 
 ID  & Period   & Detection   &   TRUFAS          & Recovered       \\  
     & (days)   &  flag       &   (FF - FS)    &    period (days)        \\  \hline
     PT &       &  1 2 3 4 5      &\\ \hline
     
 34  & 5.52     &  +++++      &      +         &  5.52     \\
 85  & 26.4     &  +++++      &      +         &  26.42     \\
 168 & 11.5     &  - - +++      &      +         &  11.51     \\
 {\it 207}& {\it 88.4}& +++++ &      -         &   -    \\
 317 &   33.8   &   - - - - -     &      -         &     -  \\
 326 &    6.8   &   - - - - -     &      -         &    -   \\
 390 &   8.0    &   +++++     &      +         &   8.00    \\
 460 &   32.9   &   +++++      &     +    &       32.93\\
 474 &  11.34   &   +++++      &     +    &      11.34 \\
 533 &   6.4    &   +++++      &     +    &       6.40\\
 537 &   2.78   &   - - + - +      &     +    &       2.78\\
 575 &   15.9   &   - - - - -      &     -    &     -  \\
 613 &    4.8   &   + - + - +      &     +    &      4.80 \\
 618 &    8.48  &   - - - - -      &    -     &      - \\
 624 &    6.7   &   + - +++      &    +     &       6.71\\
 681 &    19.8  &   - - - - -      &    -     &     -  \\
 715 &    10.1  &   - - - - -      &    -     &      - \\
 {\it 715} & {\it 63.8} &  - - - - -       &   -      &     -  \\
 835 &    42.6  & +++++        &    +     &  42.64     \\
 {\it 915}&  {\it 58.32}&  - ++ - -       &   -      &    -   \\
 917 &  30.4    &  +++++       &    +     &     30.41  \\ \hline
 EB \\ \hline
 31  & 24.7 & +++++ & + & 24.73 \\
 249  & 3.9 & +++++ & + &  3.90\\
 259  & 1.4132 & + - +++ & + & 1.41 \\
 386  & 17.1 & +++++ & + &  17.12\\
 486  & 2.4128 & - - + - + & + & 2.41 \\
 {\it 518}  &{\it 78.3}  & - - - - - & - & - \\
   599 & 1.874 & + - +++ & + & 1.88 \\
  809  &3.2  & + - +++ & + & 3.20 \\
  915 & 2.9 & +++++ & + & 2.90 \\
  919 & 13.2 & +++++ & + & 13.20 \\
   937& 8.452 & +++++ &+  & 8.45 \\
   985& 5.19 & +++++ & + & 5.19 \\ \hline
   
\end{tabular}
\end{center}
\end{table}

 In Table 2 we compare the detections in the 999 test curves as found by TRUFAS with results from the five teams that participated in BT1 (see their Tables 1 and 2), where team \# 1 used a correlation with a sliding transit template, team \# 2 used a search for box-shaped transits with lowpass filtering and broken-line detrending, team \# 3 used the BLS algorithm  (Kov\'acs et al. \cite{kovacs}) on light curves that had been detrended through a fitting of 200 harmonics, team \# 4 employed a matched filter with image-processing detrending and team \# 5 used the box-shaped transit finding algorithm by Aigrain \& Irwin (\cite{aigrain_irwin}) with an iterative 1--D filtering.  For any further details on these methods,  see BT1.

For a discussion of the detections, we should remember that BT1 included in these light curves 21 transiting planetary systems, as well as 11 low-amplitude eclipsing binaries and one eclipsing binary in a triple system; all of them were considered transit-like signals. From the 21 planets, 7 were not detected by any of the five algorithms in BT1, 9 planets where recovered by all algorithms, and 5 only by some of them. No clear ``winning algorithm'' could be identified in BT1, though team \# 3 had the best combination of detection of real events and avoidance of false detections. The performance of TRUFAS is well along the results of the best of these algorithms, except for the one case of light-curve \# 207, which was due to the limiting of the periodicity search to 50 days due to TRUFAS' requirement of at least 3 transit events.  On the other hand, TRUFAS  found {\it all} planets among those that were found only by some algorithms tested in BT1. 
As to the binary systems, one was not detected by any of the algorithms in BT1, 5 where detected by all of them, and 4 by some of them. Regarding TRUFAS performance, the same picture repeats here: it does not find periods longer than 50 days while it finds all others that were found by some of the other algorithms. The triple system (LC 249) was recovered by all algorithms including TRUFAS.

The full algorithm, from the detrending and filtering to the finding of the possible candidates,
 is implemented in IDL and it takes around 7 hours to fully process the 999 lightcurves using a PC (Dell Optiplex GX280, 1 Gb of memory and 2.8 Ghz of cpu velocity)  running under linux.  This speed corresponds to 25 seconds per light-curve. As the detrending and filtering by either of the two methods employed takes less than 2 seconds, 24 seconds can be attributed to TRUFAS itself.
 
  It is quite complicated to compare the speed of our algorithm with any of the five used in BT1. First, because this information is not given in BT1 and second, because very different parameters are involved in each of them, many with strong impact on the speed of the algorithms. However,  we have performed a more direct comparison between TRUFAS and a standard BLS method (Kov\'acs et al. \cite{kovacs}) using the 999 light curves  analysed in this paper, with the same treatment of the data and the same pre-processing  analysis. The used algorithm was the one developed by Kov\'acs et al. (\cite{kovacs}) but implemented in IDL, as the TRUFAS algorithm. Using the same period range as TRUFAS, between 1 and 60 days, the standard BLS algorithm took 43 hours to fully process the data, using 50 bins in the folded time series, as it is suggested in Kov\'acs et al. (\cite{kovacs}) as a reasonable compromise between computational efficiency and signal resolution. The number of frequencies was determined by the frequency resolution $\delta \nu = 1/(t\cdot nb)$ as 7300 (being ``{\it t}'' the total length of the series and ``{\it nb}'' the number of bins). In those conditions, the standard BLS algorithm was able to detect 21 transits. However, increasing the resolution to 60 bins and using 8760 frequencies, BLS is able to detect the 23 transits (as TRUFAS did) using 60 hours of computational time (to be compared with  7 hours spent by TRUFAS). Performing the search with an undersampling in frequencies can certainly improve the velocity of the BLS, a statement that holds for the selection of the step in the period search that is done with TRUFAS.

We note that each TRUFAS search consists of 50 trials in the peak search. Individual trials require less than one second and optimizing the numbers of trials, as well as the amount of variation between trials, might lead to significant further velocity improvements. Computation time increases approximately linearly with the number of  data-points and TRUFAS runs slightly faster if the scale on which the search is done is not selected automatically but is fixed to a typical transit-length, like the 5.7 hours that were used in our case. 

\section{Conclusions}

  In the 999 test lightcurves employed by BT1, TRUFAS provided a reliable recovery of planetary transits with periods of less than 50 days, determining the planetary periods with a precision  better than 0.15 \%. Eclipsing system's periods were recovered with a precision better than 0.3 \%. Less than 1 \% of false detections appeared when the limit of coincidence is fixed above 20 \%. All false detections were rejected  automatically with the ``A/$\sigma$'' criteria described in section \ref{sec:falsedet}.
 
 It has also been shown that this method works very well with gapped data similar to those obtained in ground-based planet search projects. Moreover, the algorithm is well suited for a completely automatic light curve processing, being fairly robust against small variations in the selection of input parameters or the kind of pre-filtering of the light curve.  Finally, TRUFAS is significantly faster than the widely employed BLS algorithm, therefore we expect that 
it may be useful in the analysis of massive transit surveys, involving ground-based projects or the upcoming satellite missions.

\begin{acknowledgements}

This work has been partially funded under grants AYA 2001-1571 and ESP 2004-03855-C03-03 of the Spanish National Research Plan. We are grateful to the authors of the ``Corot Blind Test 1'' (Moutou et al. \cite{moutou}) for providing their simulated set of stellar light-curves and to the anonymous referee, whose comments have improved the paper.
\end{acknowledgements}


\begin{thebibliography}{}

\bibitem[2004]{aigrain04} Aigrain, S., Favata, F. \& Gilmore, G. 2004, A\&A, 414, 1139

\bibitem[2004]{aigrain_irwin} Aigrain, S., \& Irwin, M. 2004, MNRAS, 350, 331

\bibitem[2003]{auvergne03} Auvergne, M., Boisnard, L. \& Buey, J.-T. 2003, SPIE 4853, 170

\bibitem[2006]{baglin06}Baglin, A. \& CoRoT Team 2006, The CoRoT Mission, ESA Special publication, in print

\bibitem[2003]{borde03} {{Bord{\'e}}, P. and {Rouan}, D. \& {L{\'e}ger}, A.} 2003, {\aap},405, 1137

\bibitem[1992]{daub} Daubechies, I. 1992, Ten Lectures of Wavelets, SIAM.

\bibitem[2006]{garrido06} Garrido, R. \&  Deeg, H.J. 2006, in Lecture Notes and Essays in Astrophysics II, Ed. A. Ulla, eConf (http://www.slac.stanford.edu/econf/), in print

\bibitem[2005]{husser} Husser, T.H., Dreizler, S., Solanki, S. \& Thomas, R. 2005, AN, 326, 628

\bibitem[2002]{jenkins} Jenkins, J.M. 2002, ApJ, 575, 493

\bibitem[2002]{kovacs} Kov\'acs, G., Zucker, S., \& Mazeh, T. 2002,  A\&A, 391, 369 

\bibitem[2005]{moutou} Moutou, C., Pont, F., Barge, P. et al., 2005,  A\&A, 437, 355 (BT1)

\bibitem[2007]{donovan} O'Donovan, F. T., Charbonneau, D., Alonso, R. et al. 2007, in print, astro-ph/0610603

\bibitem[2006]{pont06}{Pont}, F., {Zucker}, S. \& {Queloz}, D. 2006, {\mnras}, in print, {astro-ph/0608597}
\bibitem[2002]{regulo1} R\'egulo, C. \& Roca Cort\'es, T. 2002, A\&A, 396, 745

\bibitem[2005]{regulo2} R\'egulo, C. \& Roca Cort\'es, T. 2005, A\&A, 444, L5

\bibitem[2000]{rouan00} Rouan, D., Baglin, A., Barge, P. et al. 2000, ESA SP-451, 221

\bibitem[1998]{torrence} Torrence, C. \& Compo, G. P. 1998, BAMS, Vol 79, Nº 1, 61

\bibitem[1993]{young} Young, R. K. 1993, Wavelet Theory and its Applications, Kluwer Academic Publishers, p. 53-57

\end{thebibliography}
\end{document}